\documentstyle[aps,preprint]{revtex}
\begin{document}
\newcommand{\kbar}{\shortstack{\hspace*{0.02em}\_\\*[-1.0ex]\it k}}
\draft

\title{Two dimensional nonlinear dynamics  of evanescent-wave guided atoms  in
hollow fiber}
\author{X.~M.~Liu and G.~J.~Milburn}
\address{The Center for Laser Sciences, Department of Physics,}
\address{The University of Queensland, St. Lucia,
Brisbane, Qld 4072,Australia}
\date{December 9, 1999}
\maketitle

\begin{abstract}

We describe the classical and quantum two dimensional
nonlinear dynamics of large blue-detuned evanescent-wave guiding  cold
atoms in hollow fiber. We show that
chaotic dynamics exists for classic dynamics, when the intensity of the beam is
 periodically modulated.
The two dimensional distributions of atoms in $(x,y)$ plane
are simulated. We show that the atoms will accumulate on several annular
regions
when the system enters a regime of global
chaos. Our simulation shows that, when the atomic flux is very small, a
similar distribution will be obtained if we detect the atomic distribution
once each the modulation period  and integrate the signals.

For quantum dynamics, quantum collapses and revivals appear . For
periodically modulated optical potential, the variance of atomic position
will be
suppressed compared to the no modulation case.
The atomic angular momentum  will influence
the evolution of wave function in two dimensional quantum system of hollow
fiber.
\end{abstract}

\narrowtext
\section{Introduction}

Evanescent light offers a very effective means of atom manipulation because
of the
strong induced dipole interaction between atoms and evanescent field when
the field is far from resonance.
Blue-detuned evanescent light, that is light tuned above the atomic
resonance, can be used to guide atoms in a hollow
fiber\cite{Ito1},\cite {Ito2}, \cite{Renn}. With blue-detuning, atoms are
repelled from the high intensity field region
near the fiber wall. The intensity in the evanescent field is significant
over a distance
of approximately $\sim \lambda$  in the hollow region.  If the inner
diameter is several
microns, the motion of atoms will be influenced by the
evanescent field over a large transverse area inside the fiber. This system
is very suited to studying
nonlinear quantum and classical dynamics with two degrees of freedom in the
transverse plane of the fiber.
The maximum confined
transverse atomic velocities are only several centimeters per second,
therefore the atomic dynamics will be governed by the
Schr\"{o}dinger equation with an electric dipole coupling to the evanescent
field. There are few experimental studies of quantum nonlinear dynamics for
systems with two degrees of freedom and
most of our understanding of quantum nonlinear dynamics and quantum chaos
is based on simpler one degree of freedom systems,
such as an atom in a modulated standing wave\cite{Moore94}, or the
microwave ionisation of hydrogen\cite{Jensen91}. A notable
example of experimental study of quantum chaos in two degree of freedom are
the recent experiments on chaotic mesoscopc
billiards\cite{Marcus92,Taylor97}. It is clear that quite new phenomenon
can appear in two degree of freedom
systems\cite{Kovlosky94} that are absent in one degree of freedom systems.
Given the success of atom optics in providing tests
of quantum chaos in one dimension it is advisable to consider what might be
achieved using similar techniques for two
dimensional systems.   In this paper we will consider the  two dimensional
quantum and classical transverse motion of cold
atoms in a hollow fiber with a   periodically modulated evanescent field.

\section{the optical potential and  Hamiltonian  }

A two-level  atom interacting with far-off-resonant inhomogeneous laser
field has an
effective potential\cite{cohen-Tannoudji}  of the form
\begin{equation}
U({\bf r}\/)=\frac{\hbar\Delta}{2}ln(1+p),
\end{equation}
where $\Delta$ is the detuning and $p=\frac{\Omega^2/2}{\Delta^2+\Gamma^2/4}$
is a saturation parameter, with the Rabi frequency $\Omega$.
For far-off-resonant donut beams, $p \ll 1$,  and thus
\begin{equation}
U({\bf r}\/)=\frac{\hbar\Omega({\bf r}\/)^2}{4\Delta}.
\end{equation}
In experiments\cite{Ito1}, \cite{Ito2},\cite{Renn} this potential has been
used to guide cold atoms by reflecting them from
evanescent light fields on the surface of the glass. For light
striking the glass-vacuum
interface at angle $\theta$, the evanescent intensity profile is
\begin {equation}
I(r)=I(0)\alpha^2 \exp[2\kappa(r-r_{1})],
\end{equation}
where $I(0)$ is the input laser intensity at the fiber entrance, $r_1$ is the
inner radius of fiber , and the factors $\alpha$ and $\kappa$ are given in
terms of the index of refraction $n$,
inner reflection angle $\theta$ and laser wavelength $\lambda$ by
$\alpha=2\sqrt{n^2/(n^2-1)}\cos{\theta}$, and $\kappa=(2\pi/{\lambda})\sqrt{(
n^2\sin^2{\theta}-1)}$. The  Hamiltonian in transverse $(x,y)$ plane
for the system is
\begin{equation}
H_{0}=\frac{{ p^2_x+p^2_y}}{2M}+K\exp[2\kappa(r-r_{1})],
\end{equation}
where $K=\frac{\hbar\Gamma^2}{8\Delta}\frac{I(0)}{I_{s}}\alpha^2$ and
$I_s $ is the saturation intensity.

When the laser intensity  is periodically  modulated, $I(0)$ becomes time
dependent,
$I(0)[1+\epsilon\cos(\omega t)]$ and the dynamics of the atom can be
chaotic for certain initial
conditions. There are two factors to consider in treating the two
dimensional classic chaotic dynamics for evanescent wave
guided atoms in a hollow fiber. Firstly, the boundary of the hollow fiber
will limit the divergence of the atomic motion  in
the radial direction making it is easier to simulate the two-dimensional the
dynamics of the atomic system  (and also easier to probe the two dimensional
distribution in experiment).  Secondly, due to the small inner radius of  a
hollow fiber,
 atoms with a small transverse velocity will be selected at the entry to the
hollow fiber and it is not necessary to pre-cool the transverse temperature.

The effective optical potential depends on the inner radius $r_{1}$ and decay
coefficient $\kappa$. If the $r_{1}$ is much larger than wavelength $\lambda$ or
$\theta$ is too large, the potential will be very steep on the boundary and
decay
rapidly away from the inner surface of the fiber.  In order to observe two
dimensional chaotic
dynamics in $(x,y)$ plane, $r_{1}$ should be the same magnitude as
$1/\kappa$,  so that the potential will decay slowly away from the surface
of the hollow fiber.
It is possible to make the reflection angle approach the critical
reflection angle $sin^{-1}(1/n)$ by  using several
micro hollow fibers to guide the atoms and  polish the incoming end of the
hollow fiber at a specific acute angle \cite{Renn}.

We take $r_{1}=2\mu m$ and $\theta=45^0$ and consider  helium as an example.
The parameters for helium are, line width $\Gamma/{2 \pi}=1.6
MHz$, mass $M=4m_p$,  wave length  $\lambda=1.083\mu m$, saturation intensity
$I_{s}=0.16 mW/cm^2$  and recoil velocity $v_{r}=9 cm/s$.
We now define dimensionless parameters
$(\tilde{x},\tilde{y})=(2\kappa x,2\kappa y)$,
$\tilde{r}=2\kappa r$,
$( \tilde{p_x}, \tilde{p_y})=(\frac{2\kappa p_x}{M\omega_0},\frac{2\kappa
p_y}{M\omega_0})$,
 and $ \tilde{H}=H \frac{(2{\kappa}^2)}{M\omega_0^2}$,
 $\tilde{\omega}=\omega/\omega_0$
and $\tilde{t}=\omega_0 t$, where $\omega_0$ is a reference frequency.
 Omitting the tildes and defining
 $\xi=\frac{K(2\kappa)^2}{M\omega_0^2}$, the Hamiltonian
can be rewritten as
\begin{equation}
H(t)=\frac{{
p^2_x+p^2_y}}{2}+\xi e^{\sqrt{x^2+y^2}-r_{1}}(1+\epsilon\cos{\omega t}),
\end{equation}
with the canonical communication relations
\begin{equation}
 [{q_{j}}, {p_{k}}]=i\kbar\delta_{jk},
\end{equation}
where $q_j, p_k$ represents $x,y$ and $\kbar=\frac{\hbar(2\kappa)^2} {M
\omega_0}$,
plays the role of a dimensionless Planck constant.
If the dimensionless Plank constant $\kbar$ approaches $1$ ,  the atomic
motion in
small hollow fibers will be correctly described by quantum dynamics.

\section{classic chaotic dynamics and 2D distribution}

Although the flux of guided atoms entering the fiber is
very small, we still have the possibility to observe classic chaotic
dynamics by
integrating the signal over many periods of the modulation (fig. 1). The
integration length can be taken  long enough to ensure that it includes large
numbers of atoms.  We will discuss the spatial distribution of the atom
when the classical motion is chaotic.
 Using Hamilton's equations we find that the motion in
the transverse plane without modulation ($\epsilon=0$) is described by the
equations,

\begin{equation}
\dot{p_{x}}=-\xi \frac{x}{\sqrt{x^2+y^2}} e^{\sqrt{x^2+y^2}-r_{1}},
\end{equation}

\begin{equation}
\dot{p_{y}}=-\xi \frac{y}{\sqrt{x^2+y^2}} e^{\sqrt{x^2+y^2}-r_{1}}
\end{equation}

\begin{equation}
\dot{x}=p_{x},
\end{equation}

\begin{equation}
\dot{y}=p_{y}.
\end{equation}
Clearly there is only one stable  fixed point on the axis ($r=0$).

The choice of modulation frequency $\omega$ depends on the frequency
of unperturbed periodic motion. For simplicity we assume $y=0$ and
 $p_{y}=0$, so
the expression for $H$ simplifies to a one dimensional Hamiltonian system.
The period of motion  for the unperturbed Hamiltonian $H_{0}$ is\cite{lich}
\begin{equation}
T_{0}=\oint{\frac{dx}{\partial H_{0}/\partial p_{x}}
= 2 \int_{-x_M}^{x_M} \frac{dx}{\sqrt{2(H_{0}-\xi e^{\sqrt{x^2}-r_{1}})}}},
\end{equation}
where $x_{M}$ is determined by $H_{0}= \xi e^{\sqrt{x^2_{M}}-r_{1}}$.
Therefore
\begin{equation}
\omega_{0}=\frac{\pi}{\int_{-x_M}^{x_M} \{2[H_{0}-\xi e^{\sqrt{x^2}-r_{1}}]\}
^{-1/2}dx}.
\end{equation}
The graph of $\omega_{0}$ versus $H_{0}$ and $ x_{M}$ versus $H_{0}$ is shown
in fig. 2. We can select the modulation frequency
$\omega$ to control the  position of the fixed points.
Here we take $\omega_0=265 KHz$ ,  dimensionless modulation frequency
$\omega=2$
and set $\xi=50$. The new fixed points will appear in $x \approx 0.41$ and 
$x \approx 2.87$.

We use a symplectic integration routine \cite{forest},\cite{dyrting} to
solve the equations
of motion so as to preserve the Poisson bracket
relation $\{x(t),p_x(t)\}=1$, and thus maintain the Hamiltonian character
of the motion.
We  plot the stroboscopic portrait of
the system at multiples of the period of modulation, $t=(2\pi/\omega)s$,
where s is an integer referred to
as the strobe
number. From fig. 3 we can see that regions of globally chaotic motion will
arise  when $\epsilon$ is large, together with some regular
regions. A broad initial phase space distribution of atoms will
enable  some atoms to become trapped in these stable regions.

Laser cooling and trapping techniques have the ability to cool the atom to
very low velocities and trap them with  well localized momentum, however
the position distribution is not so well localized.
Therefore the appropriate description of the initial conditions is in terms
of a probability density on phase
space $(x,y,p_x,p_y)$. We define a classical state to be a probability measure
on phase space of the form $Q(x,y,p_x,p_y)dxdydp_xdp_y$. The probability
density satisfies the Liouville equation
\begin{equation}
\frac{\partial{Q}}{\partial{t}}={\{H, Q\}}_{q_i,p_i},
\end{equation}
where${\{,\}}_{q_i,p_i}$ is the Poisson bracket. This equation can be solved by
the method of characteristics.
To simulate an experiment, we assume atoms
are initially randomly uniformly distributed on $x^2+y^2 < {r_{1}}^2$,

The momentum distributions for $p_x$ and $p_y$ are assumed to be Gaussian
distributions. Therefore
\begin{equation}
Q_{0}(x,y,p_x,p_y)=Q_{0}(x)Q_{0}(y)Q_{0}(p_x)Q_{0}(p_y),
\end{equation}
where
\begin{equation}
Q_{0}(p_i)=\frac{1}{2\pi\sigma_{p_{i}}}\exp{[-{(p_i-p_i(0))^2}/{2\sigma_{p_{i}}]
}}.
\end{equation}
The variances of $p_x$ and $p_y$ are related to the temperature $T_{i}$
\begin{equation}
\sigma_{p_{i}}=k_{B} T_{i}/[M\omega_s^2/(2\kappa)^2].
\end{equation}

We simulated the atomic system of $10^4$ numbers and  take
$\sigma_{p_{i}}=0.1$,
which corresponds to radial rms velocity $2cm/s$ for helium if $
\theta=45^0$ and
$\omega=2$ .
The variance of $Q$ function with time is
\begin {equation}
Q(\bf{r}, \bf{p}, t)=Q_{0}[ \bar{\bf{r}}(\bf{r}, \bf{p}, -t),
\bar{\bf{p}}(\bf{r}, \bf{p}, -t)]
\end{equation}

In the case of no modulation, atoms will accumulate
around the fix point $x=y=0$. When the modulation is added the
atoms will
diffuse in regions of chaotic motion but some will accumulate around
several rings corresponding to fixed points at non-zero
radius (fig. 4).

Because the inner size of hollow fiber is very small. The flux of guiding
atoms
 entering the fiber will be low. But we can use a large blue-detuned 
 shield laser,
 with the duration period of square wave   equals to the modulation period
 $T={2\pi}/{\omega}$,
 and the time interval $ dT << T$. When atoms emit out from the hollow fiber,
 they will travel freely without the optical potential. If the shield
 laser turn on, it will block atoms enter the detection region. In the short
 time interval $ dT $, atoms will enter the detection region and the detector
 records the atomic distribution of momenta and positions. This procedure can be
 repeated as long as possible till we have integrated enough atomic numbers for
 many snap shots. The simulation shows that for integration of atomic numbers at
 different integer strobe numbers ,  it will produce similar results (fig. 5).

 Because rings (fig. 4)represent new fixed points in
phase space and the atomic radial momenta are almost approach zero(fig. 3).
When atoms emit out and travel freely the shape of rings will not change very 
much in 
detection region. But the  atomic velocity orientations outside the fixed 
points are  randomly distributed.
In order to measure the rings and small spatial distribution of atoms, 
high precision position
measurements are required. The Raman-induced resonance imaging method 
\cite{Thomas},\cite{Stokes}, \cite{Gardner} can be used to measure the
positions with nanometer spatial resolution limited by the quantum uncertainty
principle.

\section{Two dimensional quantum nonlinear dynamics}

If the transverse temperature of atoms in hollow fiber is very cold,
quantum nonlinear dynamics will result. Here we take the same parameters
and same definition for dimensionless parameters
as the classical dynamics and the dimensionless Plank constant
$\frac{\hbar(2\kappa)^2} {M \omega_s} \simeq 1$.
 The dimensionless 2D Schr\"{o}dinger equation for atoms in hollow fiber
\begin{equation}
i\kbar\frac{\partial\psi({\bf r}\/,t)}{\partial t}=\hat{H}(t)\psi({\bf r}\/,t),
\end{equation}
where
\begin{equation}
\hat{H}(t)=-\frac{\kbar^2}{2}(\frac{\partial^2}{\partial
x^2}+\frac{\partial^2}{\partial
y^2})+V(x, y,t).
\end{equation}
and
\begin{equation}
V(x, y, t)=\xi e^{\sqrt{x^2+y^2}-r_{1}}(1+\epsilon\cos{\omega t}),
\end{equation}
where $x^2+y^2 \leq {r_{1}}^2$.

For simplicity we assume the boundary condition for eq. (18) is
\begin{equation}
\psi(x, y, t)|_{\sqrt{x^2+y^2} = r_{1}}=0.
\end{equation}

  The Split Operator Method \cite{Kosloff}and 2D FFT (Fast Fourier
Transformation) \cite{Press} will be used to obtain the numerical solution
of Schr\"{o}dinger
eq. 18. The scheme in which the kinetic operator and potential operator
are used to propagate the wave function separately:
\begin {equation}
\exp[-i\hat{\bf H}\/{\delta}t/\kbar]\sim \exp[-i(\hat{\bf
P}\/)^2{\delta}t/{4\kbar}]
\exp[-i(\hat{\bf
V}\/){\delta}t/{2\kbar}]
\exp[-i(\hat{\bf
P}\/)^2{\delta}t/{4\kbar}],
\end{equation}
and the computing errors are of $O(\delta t^3)$.

We assume  at $t=0$ the wave function is a  minimum  uncertainty wave
function. The initial variance of $x, y$ are the
same $\sigma_{x}=\sigma_{y}=\sigma$. The expression for
this wave function is
\begin{equation}
\psi(x, y) =\frac{1}{\sqrt{2\pi\sigma}}e^{-\frac{(x-x_{0})^2}{4\sigma}}
e^{-\frac{(y-y_{0})^2}{4\sigma}}e^{iP_{0x}x/\kbar}e^{iP_{0y}y/\kbar}.
\end{equation}

In order to observe genuine quantum nonlinear behavior we have to wait for a
time that is longer than the classical period. The time scale for quantum
nonlinear dynamics $T_{rev}$ is given by expression\cite{chen}
\begin{equation}
T_{rev}=T_{0} (\frac{\kbar}{2}|\partial \omega_{0}/\partial{\bar{E}}|)^{-1}.
\end{equation}
In fig. 6 we have plotted the momentum mean $<p_{x}>$ and its variance as a
function of time for $\epsilon=0$. Quantum collapses and revivals appear for
quantum nonlinear dynamics as expected.

In order to understand the influence of the modulation of the potential, we
have plotted the variances
of  $<x^2>-<x>^2$, $<{p_{x}}^2>-<{p_{x}}>^2$ versus strobe numbers.
In fig. 7 we found at integer
 strobe numbers,  the modulation will increase the
 variance of momentum , but suppress the variance of position.
 The fluctuation of variances of positions and momentum are consistent with the
 Heisenberg uncertainty relationship in this quantum system. 

If we rewrite $\hat{H}$ in polar coordinator$(r, \theta)$, we get
\begin{equation}
\hat{H}=-\frac{\kbar^2}{2}(\frac{\partial^2}{\partial
r^2}+\frac{1}{r}\frac{\partial}{\partial
r}+\frac{1}{r^2}\frac{\partial^2}{\partial\theta^2})+V(r, t),
\end{equation}
The general solution of eq.(19) will be \cite{Collatz}
\begin {equation}
\psi(r, \theta, t)=\sum_{m}{{a_{m} (r, t)} e^{ im\phi}},
\end {equation}
where m is angular momentum quantum number. It shows that the evolution of
wave function will be related to atomic angular
momentum. In general  for $t > 0$ the probability ${|\psi(x, y)|}^2$ will not
be symmetric  except the initial wave function has no  angular
momentum. In fig. 8 for initial conditions $x_{0}=y_{0}=0$ and
$p_{0x}=p_{0y}=0$ we have
plotted th the probability at $y=0$ plane
as a function of $x$ at the time of  strobe number $50$, it shows that it
will stay
symmetric both for $\epsilon =0$ and $0.7$. In fig. 9, for initial condition
$x_{0}=y_{0}=0$ and $p_{0x}=p_{0y}=1$ , the probability
distribution for $t > 0$ will  not be symmetric because
initial atomic wave functions include angular momenta, and the interaction
between atomic momenta and optical
potential will destroy the spatial symmetry of probability. Therefore the
atomic
angular momentum plays an important rule in the evolution of wave function.

\section{conclusion and discussion}

The atomic motion in small diameter hollow fiber is quantum mechanical if
dimensionless Plank constant $\kbar$ approach $1$ .
Quantum collapses and revivals will appear in the dynamics of the mean
values . If
the intensity is periodically modulated, we found the modulation will increase
the variance of momentum , but suppress the variance of position
at integer  strobe numbers.  The fluctuation of variances of positions and
momentum
is consistent with the Heisenberg uncertainty relationship in this  system.
In  this two degree of freedom quantum system,  the atomic angular momentum
will influence the evolution of wave function.

Although the flux of guiding atoms entering the fiber is
low, it's still possible to observe the classic chaotic
dynamics by
integrating the signals at different strobe numbers in experiment. The
integration length can be taken  long enough to ensure that it includes a large
numbers of atoms. We have shown that an atom moving in an intensity modulated
evanescent-wave field in hollow fiber can exhibit chaotic dynamics in the
transverse plane. For atomic momenta $p_{x}$, $p_{y}$ with Gaussian
distributions, some atoms will become trapped in rings
corresponding to radial fixed points of the modulated system in the moments of
integer strobe numbers.

 For the atomic average radial velocity is around recoil velocity and the spatial
 distribution size no larger than inner diameter of fiber, high resolution 
 position and velocity measurements must be used.The Raman-induced resonance 
 imaging method 
can reach the  nanometer spatial resolution limited by the uncertainty 
principle.
For atomic momentum distribution can be measured by Time of Flight absorption
imaging method{\cite{Davis}, \cite{Bradley}} which is  used to measure the
temperature of super cold atoms in BEC experiment, or atomic velocity selection
method using stimulated Raman transitions\cite{Kasevich}.

It is quite interesting that both classical chaotic dynamics and quantum
dynamics have the possibility to be realized in experiment for atoms 
propagating in hollow fiber. Because the fiber's inner size is very small,
it will select atoms with very
small radial velocity to enter the fiber and there is no need to pre-cool
the radial
motion. In addition the fiber's boundary will limit the  divergence  of
radial motion, making  it is easier to observe both two dimensional 
classical and quantum dynamics in experiment.  We believe this is a far
more practical scheme for observing the classical and quantum
nonlinear dynamics of radially confined atoms than other schemes such as
those that use a donut beam scheme\cite{liu}.

\section{Acknowledgment}
One of authors(XML) would like to thank
Dr. Kenneth Baldwin at Australia National University for useful discussion about
physical parameters of helium and hollow fiber.

\newpage

\begin{figure}[htbp]
\caption{{\em The diagram of proposed experiment. The large blue-detuned shield
laser will block atoms enter the detection region except in the short time
interval $ dT << T$. The integration of many snap shots will have the similar
effect of atomic distribution compared to one snap shot of large atomic numbers.
}}
\protect\label{fig_NM}
\end{figure}

\begin{figure}[htbp]
\caption{{\em  a. The relations between frequency of motion and Hamiltonian
$\omega_{0} \sim H_{0} $ and $x_{M}\sim H_{0}$.} }
\protect\label{fig_NM}
\end{figure}

\begin{figure}[htbp]
\caption{{\em  Stroboscopic portrait of the system with $\epsilon=0.7$,
$p_x(0)=0$, $p_{y}=0$ and $y=0$. The maximum strobe number is $500$.} }
\protect\label{fig_NM}
\end{figure}

\begin{figure}[htbp]
\caption{{\em  The atomic distribution in $(x,y)$ plane at the strobe
number 50
for $\epsilon=0.7$. The $10^4$ atoms were taken in phase space. The atoms
were initially
distributed on $x^2+y^2 \leq r^2_0$ region. The momenta
of $p_x$, $p_{y}$ are Gaussian distributions and
$\sigma_{p_{x}}=\sigma_{p_{y}}=0.1$..} }
\protect\label{fig_NM}
\end{figure}

\begin{figure}[htbp]
\caption{{\em The atomic distributions of $10^4$ numbers in $(x,y)$ plane.
After strobe number 50,
we count 200 atoms at the time of every integer strobe number until
$s=150$. } }
\protect\label{fig_NM}
\end{figure}

\begin{figure}[htbp]
\caption{{\em The average momentum $<p_{x}>$ and momentum variance
$<{p_{x}^2}>-{<p_{x}>}^2$ evolutions with time. The initial wave function
is the least uncertainty state, $\sigma_{x}=\sigma_{y}=0.1$. For (a),
  $x_{0}=y_{0}=0$, $p_{0x}=p_{0y}=1.0 $. For (b),
  $x_{0}=y_{0}=0.5$, $p_{0x}=p_{0y}=0.0 $.} }
\protect\label{fig_NM}
\end{figure}

\begin{figure}[htbp]
\caption{{\em The position variance$<{x}^2>-<x>^2$ and momentum variance
$<{p_{x}^2}>-{<p_{x}>}^2$ versus strobe numbers. The initial wave function
is the least uncertainty state, $\sigma_{x}=\sigma_{y}=0.1$. For (a) and (b)
  $x_{0}=y_{0}=0$, $p_{0x}=p_{0y}=1.0 $. For (c) and (d),
  $x_{0}=y_{0}=0.5$, $p_{0x}=p_{0y}=0.0 $.} }
\protect\label{fig_NM}
\end{figure}

\begin{figure}[htbp]
\caption{{\em The probability  distributions ${|\psi(x, 0)|}^2 $ for $y=0$.
The initial wave function
is the least uncertainty state, $\sigma_{x}=\sigma_{y}=0.1$,  $x_{0}=y_{0}=0$,
$p_{0x}=p_{0y}=0$.The point line is distribution for $t=0$, the dashed line is
distribution for $t=50 T$ ,  no modulation $\epsilon=0$, and the solid line
is the
distribution for $t=50 T$ ,  $\epsilon=0.7$. } }
\protect\label{fig_NM}
\end{figure}

\begin{figure}[htbp]
\caption{{\em The probability  distributions ${|\psi(x, 0)|}^2 $ for $y=0$.
The initial wave function
is the least uncertainty state, $\sigma_{x}=\sigma_{y}=0.1$,  $x_{0}=y_{0}=0$,
$p_{0x}=p_{0y}=0.1$. The point line is distribution for $t=0$, the dashed
line is
distribution for $t=50 T$ , no modulation $\epsilon=0$, and the solid line
is the
distribution for $t=50 T$ , $\epsilon=0.7$.} }
\protect\label{fig_NM}
\end{figure}

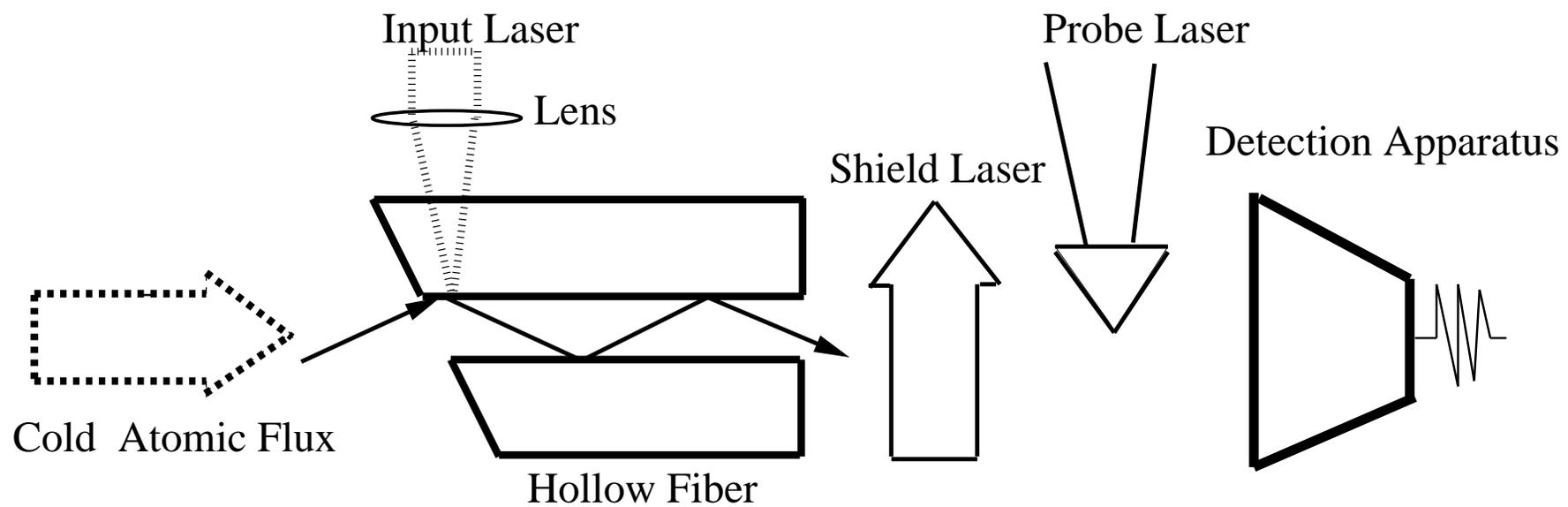
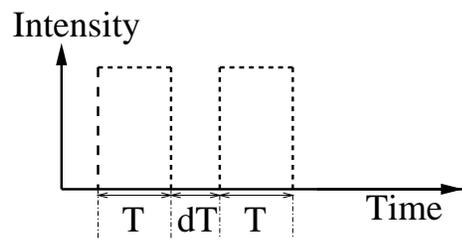
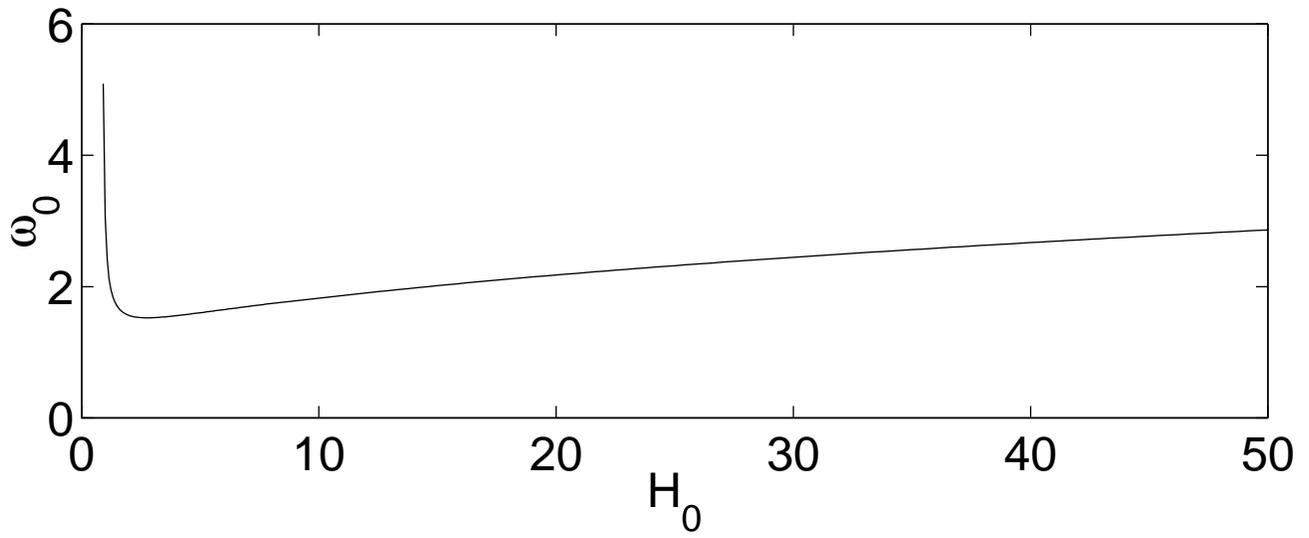
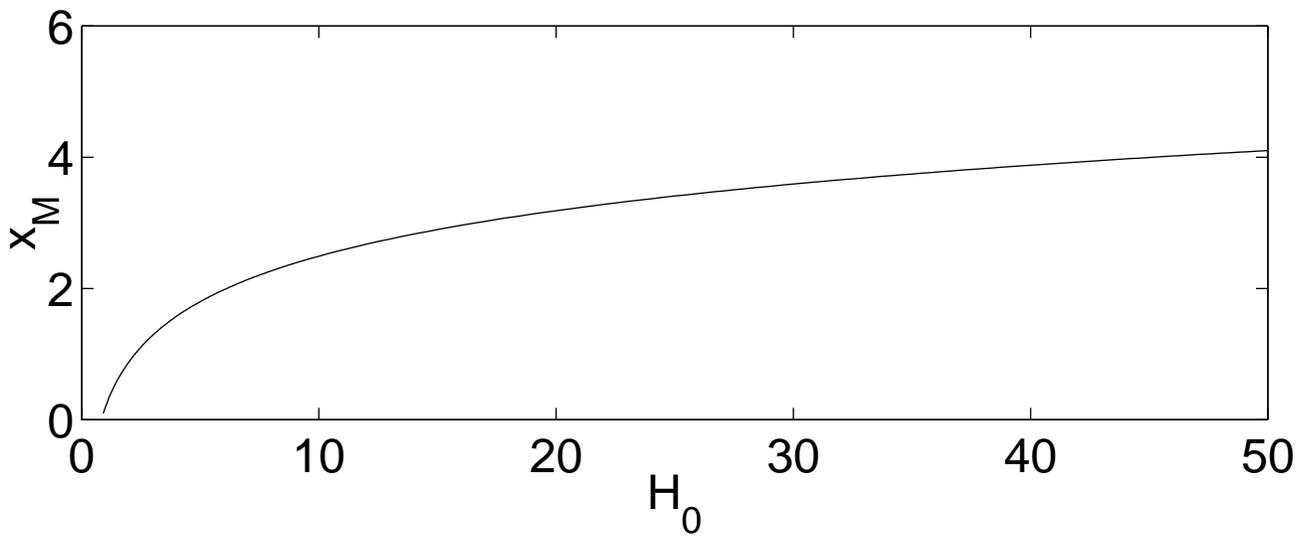
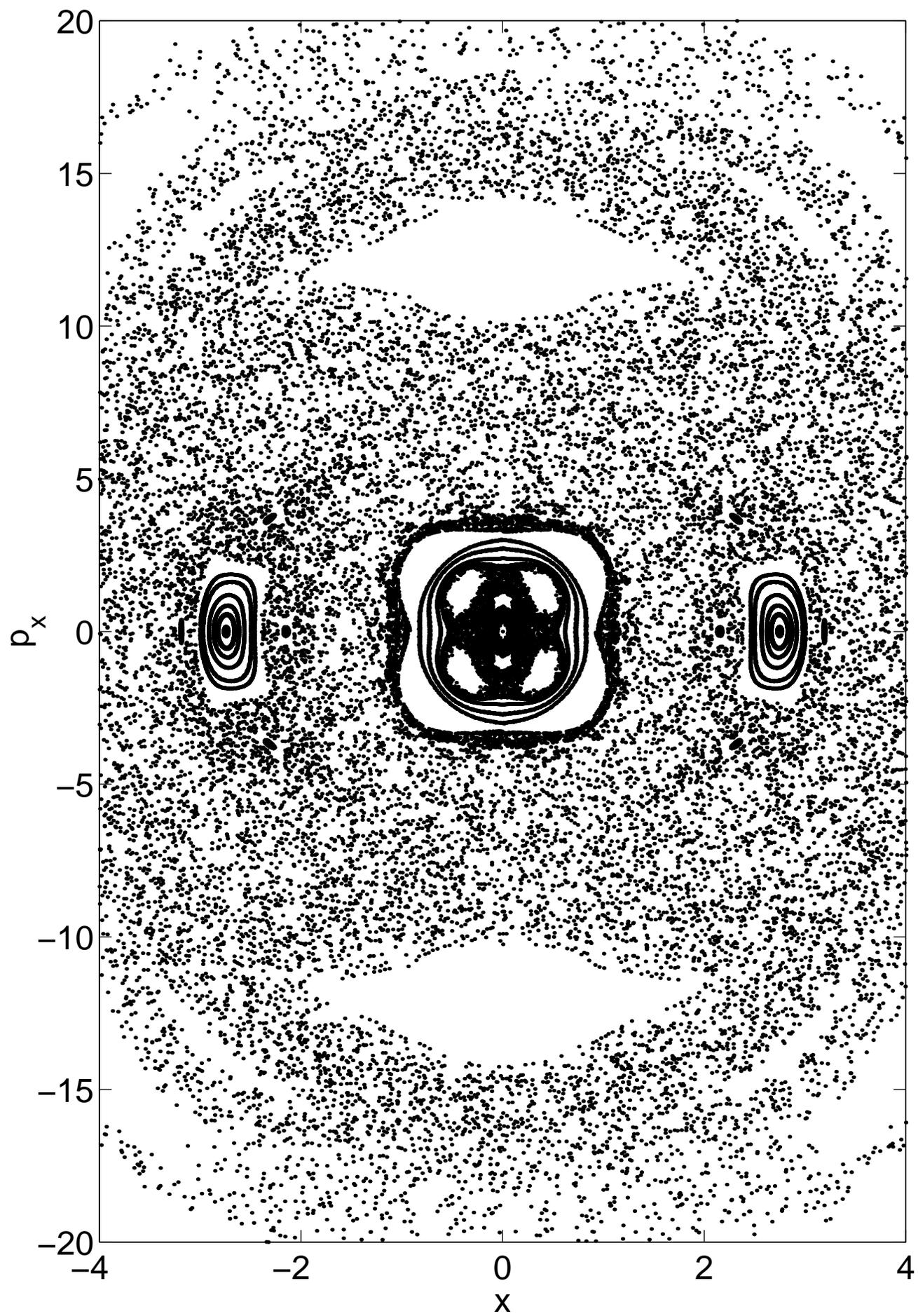
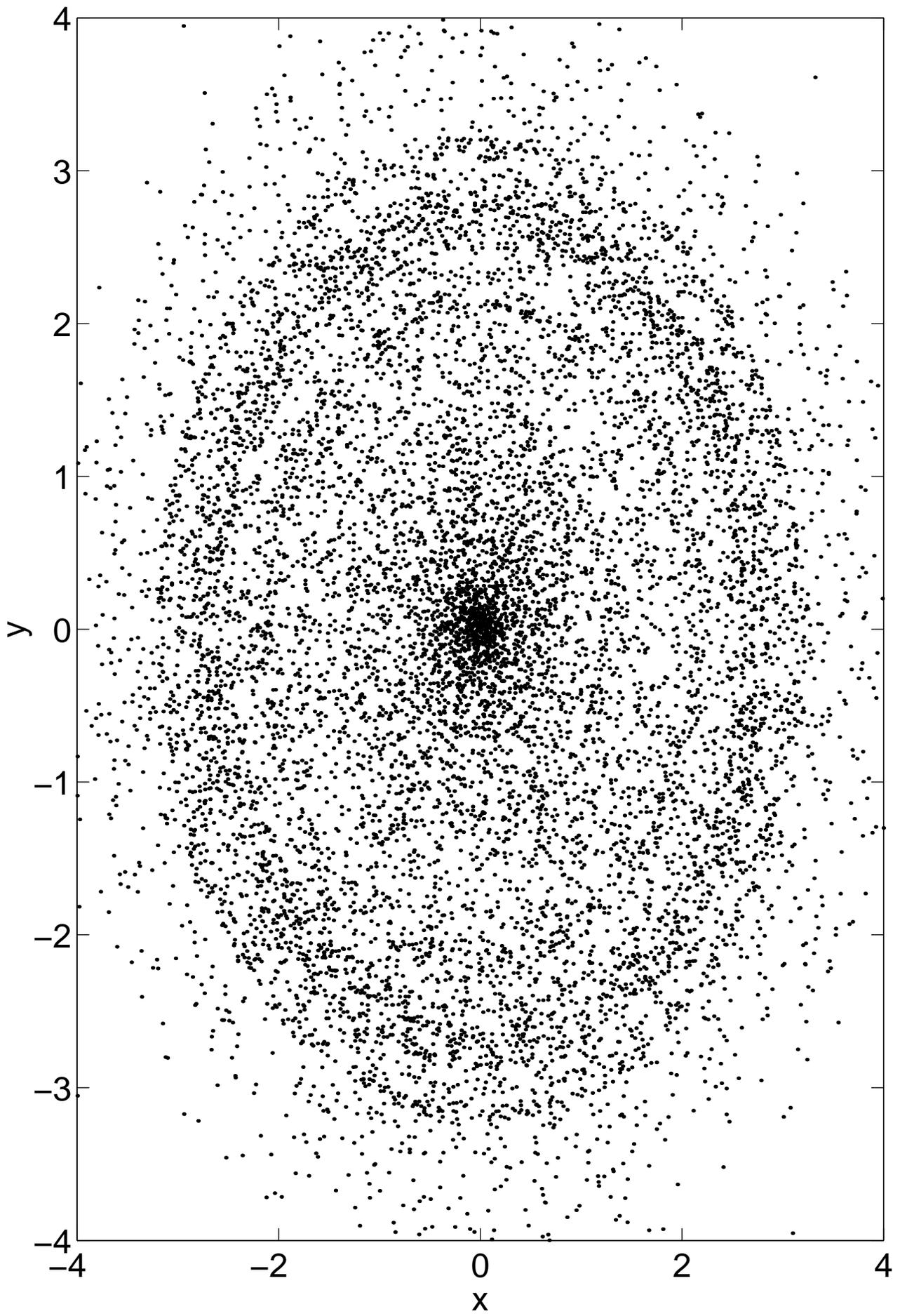
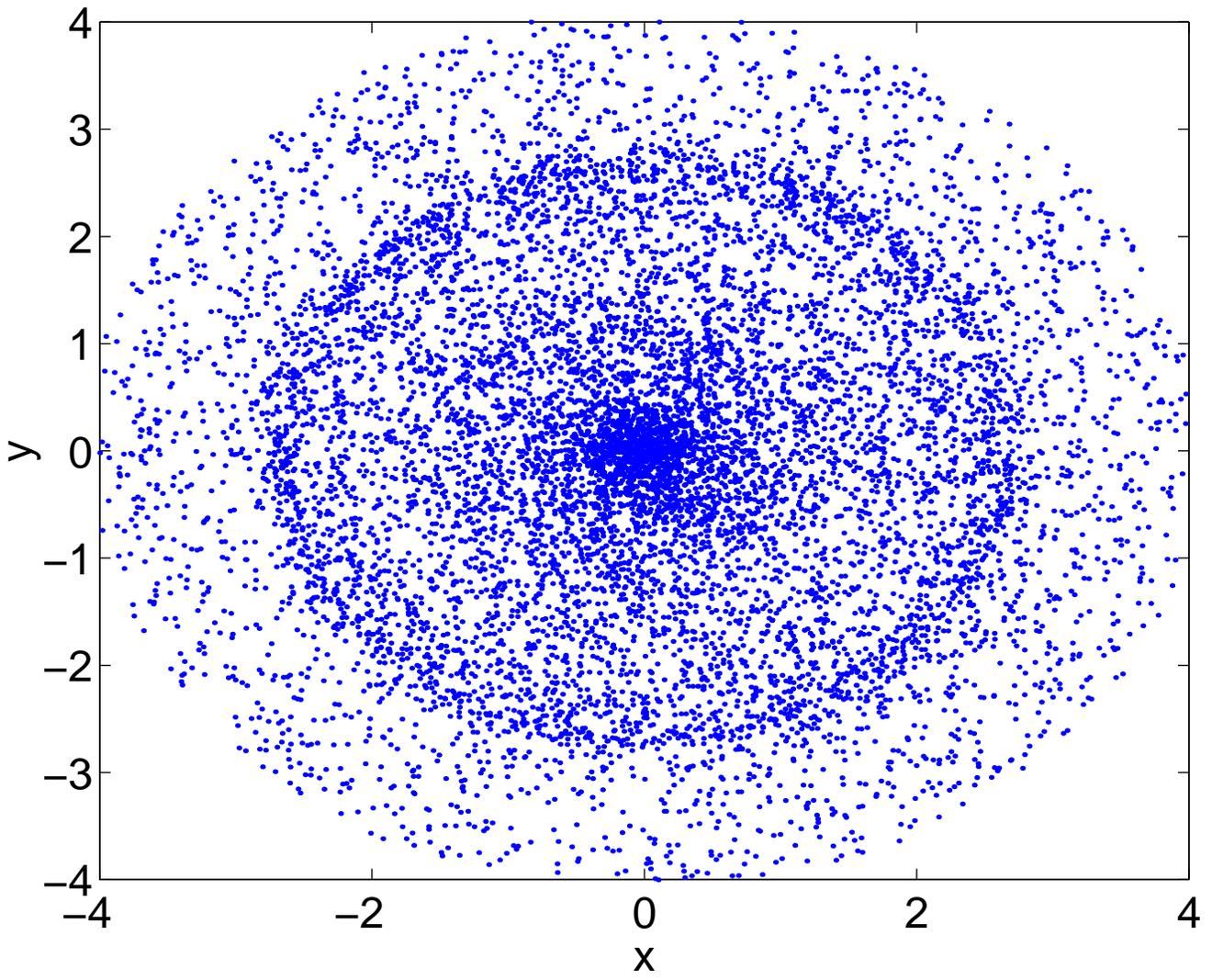
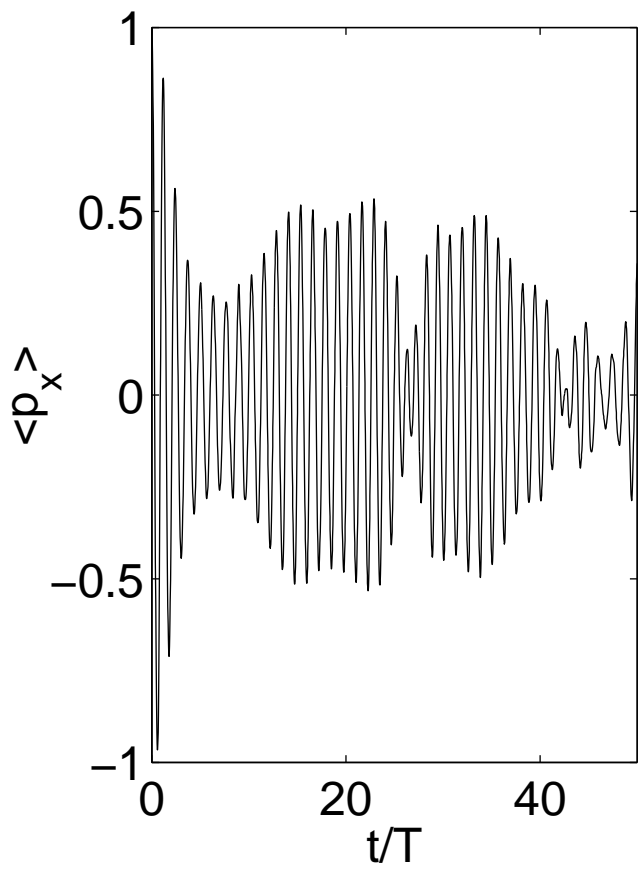
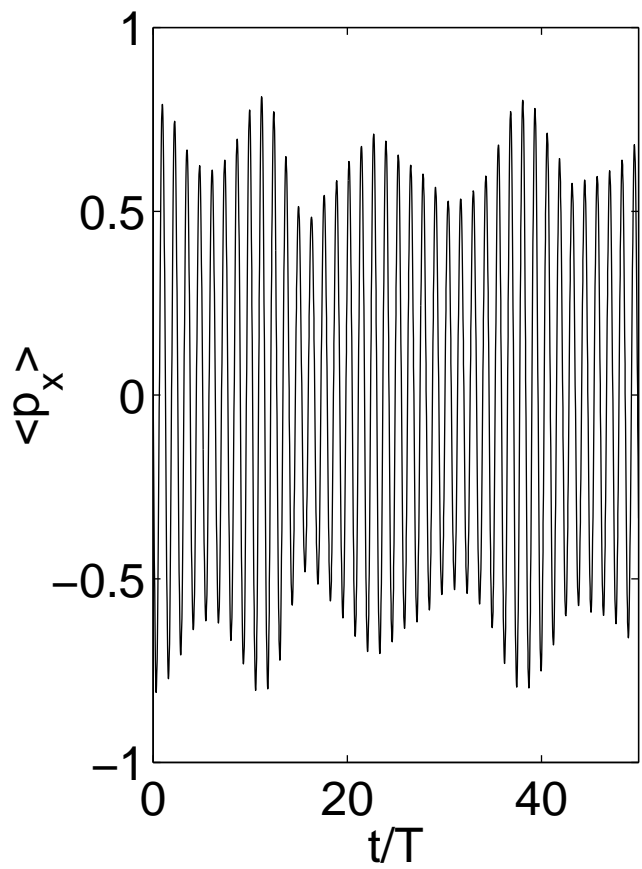
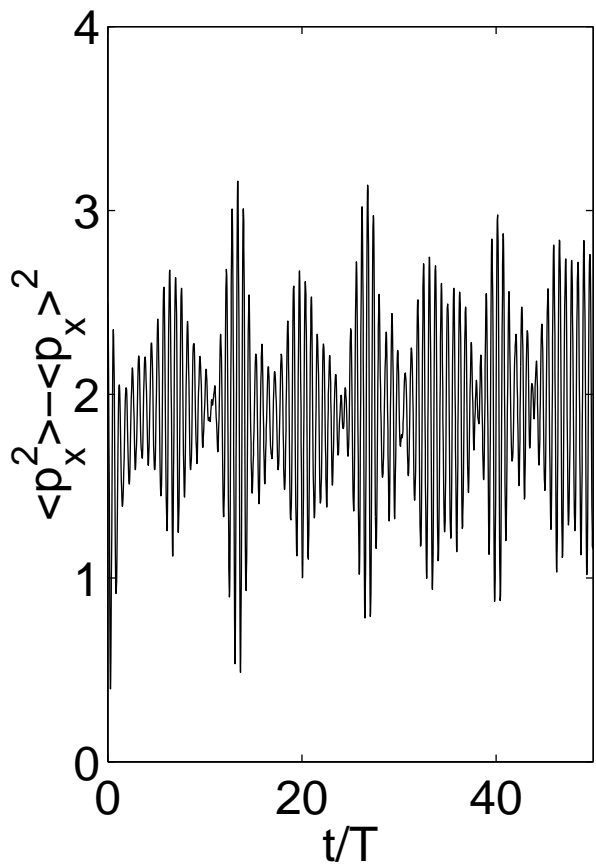
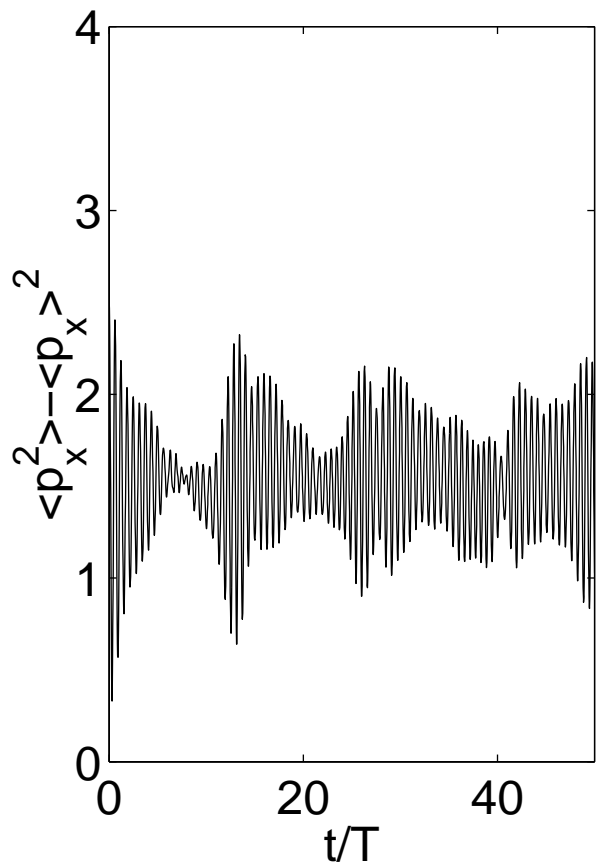
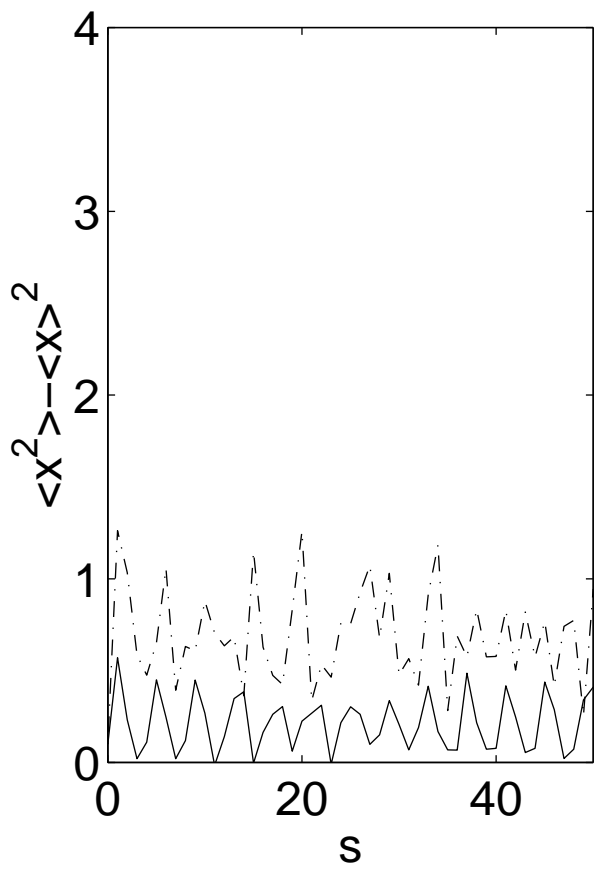
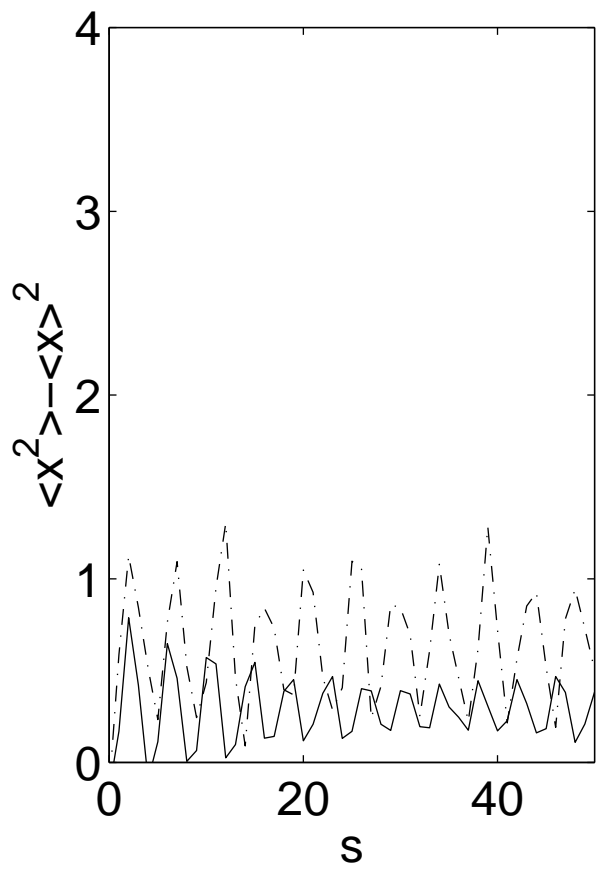
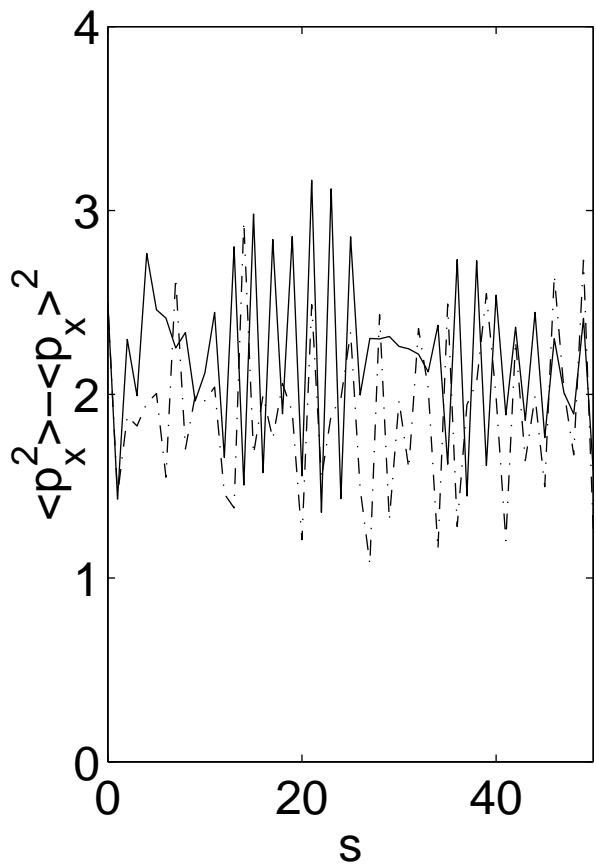
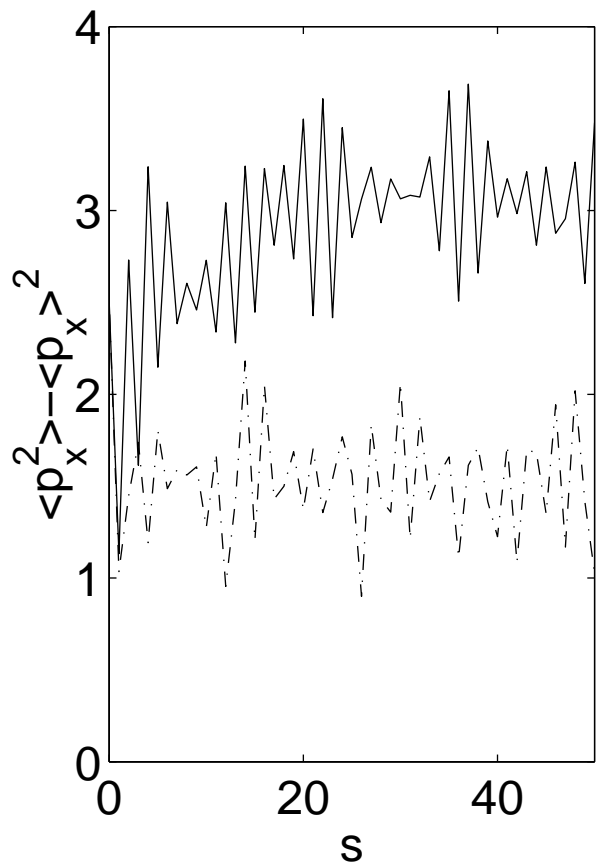
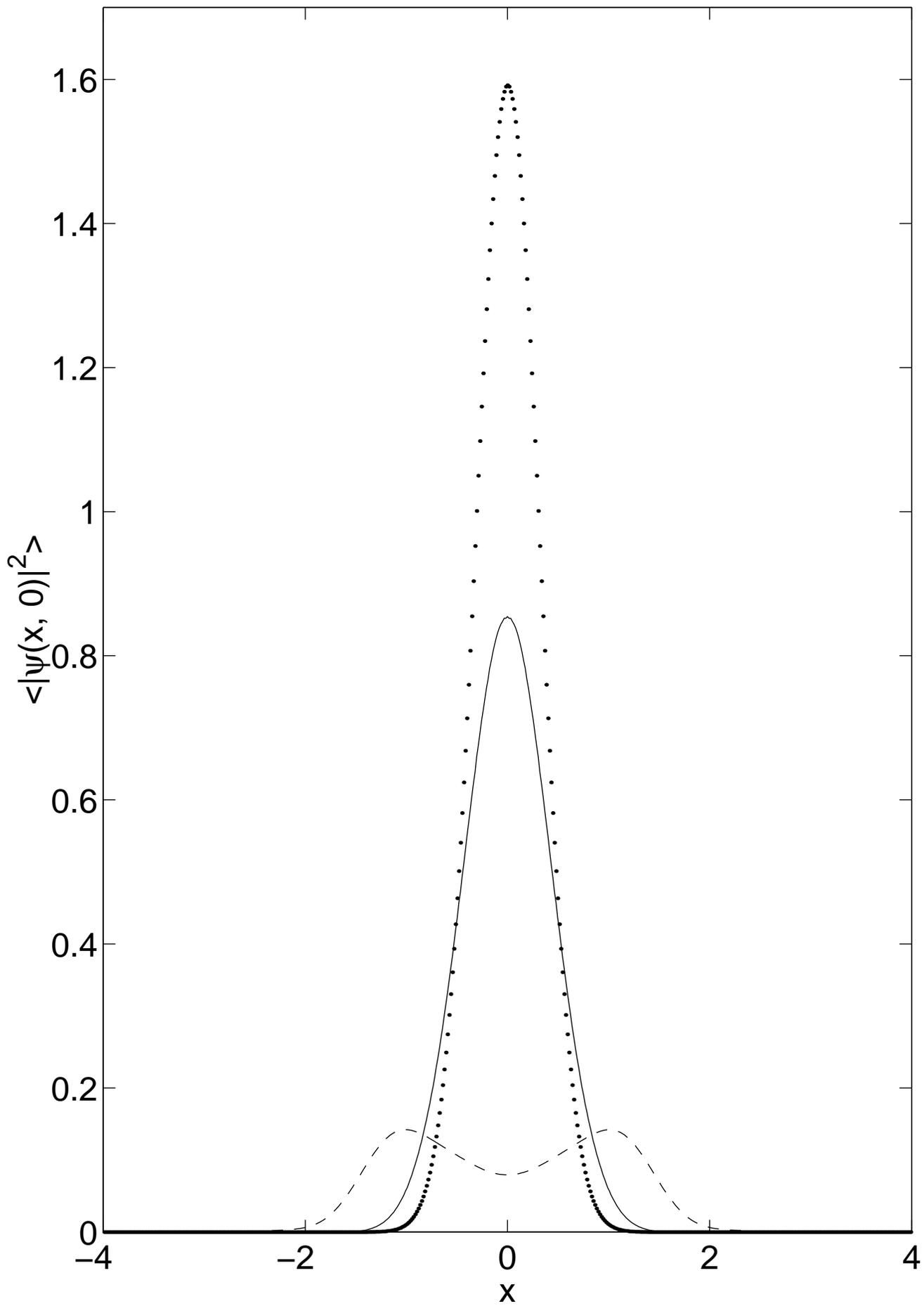
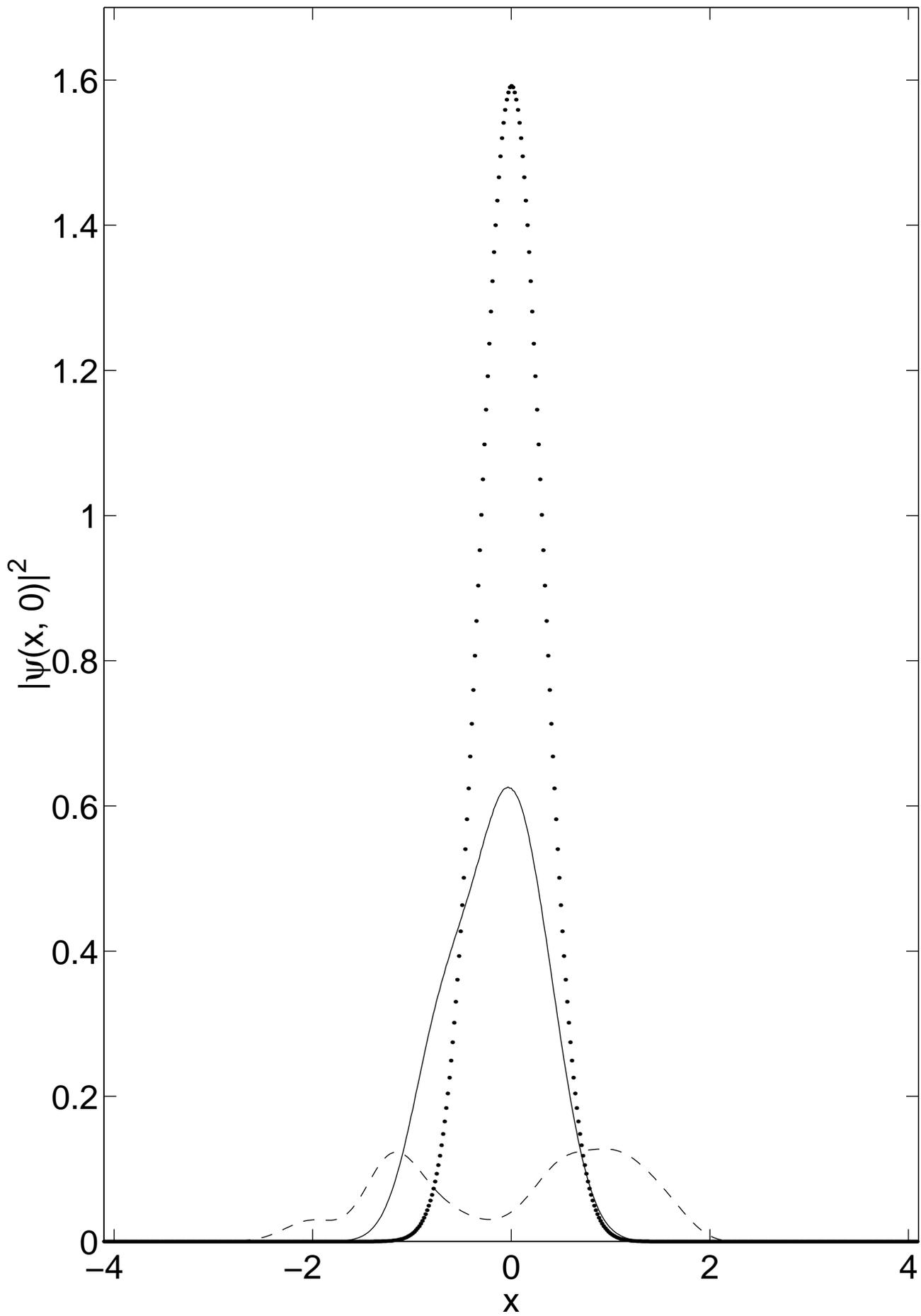
\end{document}